\documentclass[aps,prd,reprint,superscriptaddress,amsmath,amssymb,nofootinbib,nolongbibliography,floatfix]{revtex4-2}
\usepackage{graphicx}
\usepackage{dcolumn}
\usepackage[utf8]{inputenc}
\usepackage{xcolor}
\usepackage[unicode=true,colorlinks,linkcolor=blue,anchorcolor=blue,urlcolor=blue,citecolor=blue,breaklinks=true]{hyperref}
\usepackage{subfigure}

\newcommand{\ii}{\mathrm{i}\,}

\newcommand{\pararrow}{\mathord{\buildrel{\lower3pt\hbox{$\scriptscriptstyle\leftrightarrow$}}\over {\partial}}} 
\newcommand{\pararrowk}[1]{\mathord{\buildrel{\lower3pt\hbox{$\scriptscriptstyle\leftrightarrow$}}\over {\partial}\hspace*{-0.18em}{}^#1}\hspace*{-0.18em} \,} 

\usepackage{physics}
\usepackage{bm}

\newcommand{\qfnu}{\affiliation{College of Physics and Engineering, Qufu Normal University, Qufu 273165, China}}

\newcommand{\itp}{\affiliation{CAS Key Laboratory of Theoretical Physics, Institute of Theoretical Physics, Chinese Academy of Sciences, Beijing 100190, China}}

\newcommand{\imp}{\affiliation{Institute of Modern Physics, Chinese Academy of Sciences, Lanzhou 730000, China}}

\newcommand{\snst}{\affiliation{School of Nuclear Science and Technology, University of Chinese Academy of Sciences, Beijing 101408, China}}

\newcommand{\scnt}{\affiliation{Southern Center for Nuclear-Science Theory (SCNT), Institute of Modern Physics, Chinese Academy of Sciences, Huizhou 516000, Guangdong, China}}

\begin{document}

\title{Charmless decays of the spin-2 partner of $X(3872)$}

    \author{Zu-Xin Cai} \qfnu \imp
	\author{Zhao-Sai Jia} \qfnu\itp
	\author{Gang Li}\email{gli@qfnu.edu.cn} \qfnu
	\author{Shi-Dong Liu}\email{liusd@qfnu.edu.cn} \qfnu 
	\author{Ju-Jun Xie} \email{xiejujun@impcas.ac.cn} \imp \snst \scnt

\begin{abstract}

The Belle collaboration recently reported a promising candidate for the spin-2 $D^*\bar{D}^*$ partner of the $X(3872)$, called the $X_2$ for short, having a mass of $(4014.3 \pm 4.0 \pm 1.5)~\mathrm{MeV}$ and a width of $(4 \pm 11 \pm 6)~\mathrm{MeV} $. In present work, we assume the $X_2$ as a pure molecule of the $D^*\bar{D}^*$ under three cases, i.e., pure neutral components ($\theta = 0$), isospin singlet ($\theta = \pi/4$) and neutral components dominant ($\theta = \pi/6$), where $\theta$ is a phase angle describing the proportion of neutral and charged constituents. Using an effective Lagrangian approach, we calculated the partial widths of $X_2\to VV$ and $X_2 \to PP$ ($V$ and $P$ stand for light vector and pseudoscalar mesons, respectively). The predicted decay widths of $X_2 \to VV$ can reach a few hundreds of $\mathrm{keV}$, while the decay widths of $X_2 \to PP$ are about several tens of $\mathrm{keV}$. In addition, the effects from the proportion of neutral and charged constituent on the decay widths of $X_2\to VV$ and $PP$ are also investigated. We hope that the present calculations will be checked experimentally in the future.

\end{abstract}

\date{\today}

\maketitle


\section{Introduction} \label{section:introduction}

Since the Belle Collaboration first reported the observation of the $X(3872)$ in the $\pi^+\pi^-J/\psi$ invariant mass spectrum from the $B \to K \pi^+ \pi^- J/\psi$ decay in 2003~\cite{Belle:2003nnu}, there has been a surge of interest in the field of exotic state research. In 2013, the LHCb Collaboration established the quantum numbers of the $X(3872)$ to be $J^{PC} = 1^{++}$~\cite{LHCb:2013kgk}. Currently, the $X(3872)$ has a world average mass of $(3871.65\pm 0.06)~\mathrm{MeV}$ and an exceptionally narrow full width of $(1.19\pm 0.21)~\mathrm{MeV}$~\cite{2022pdgPoTaEP2022-083C01}. The mass of $X(3872)$ is extremely close to the $D^0 \bar{D}^{*0}$ threshold ($m_{D^0} + m_{{\bar D}^{*0}} = 3871.69~\mathrm{MeV}$), thus it leads to the natural explanation of the $X(3872)$ as a $D \bar{D}^{*}$ hadronic molecule~\cite{Close:2003sg,Pakvasa:2003ea,Swanson:2004pp,Swanson:2003tb,Tornqvist:2004qy,Voloshin:2003nt,Wong:2003xk,AlFiky:2005jd,Braaten:2006sy,Fleming:2007rp,Ding:2009vj,Dong:2009yp,Lee:2009hy,Lee:2011rka,Liu:2009qhy,Zhang:2009vs,Gamermann:2009uq,Mehen:2011ds,Nieves:2011zz,Nieves:2012tt,Li:2012cs,Sun:2012sy,Guo:2013sya,He:2014nya,Zhao:2014gqa,Guo:2014taa,Guo:2014hqa,Braaten:2003he,Fleming:2008yn,Meng:2021kmi,Wu:2021udi,wang2014EPJC74-2891}. The comprehensive molecular interpretation of the $X(3872)$ can be found in the reviews~\cite{Guo:2017jvc, Kalashnikova:2018vkv}. Other interpretations, e.g., the compact tetraquark \cite{Wang:2023sii,Wang:2013vex} and a conventional charmonium state \cite{Barnes:2003vb,Suzuki:2005ha} are also possible.

If the $X(3872)$ is a mesonic molecule of the $D\bar{D}^*$ with $J^{PC} = 1^{++}$, there would exist of a bound state of the $D^* \bar{D}^{*}$, the spin-2 partner of the $X(3872)$, based on the heavy quark spin symmetry (HQSS), which is usually called $X_2$ with the quantum numbers $I^{G}(J^{PC}) = 0^{+}(2^{++})$~\cite{Guo:2014ura,Albaladejo:2015dsa,Nieves:2012tt,Hidalgo-Duque:2012rqv,Hidalgo-Duque:2013pva,Baru:2016iwj,Shi:2023mer}.
The predicted mass of the $X_2$ is around $4012~\mathrm{MeV}$, with a binding energy and a width similar to those of the $X(3872)$. Subsequently, considerable theoretical work was conducted to investigate the $X_2$ from various perspectives~\cite{Hidalgo-Duque:2012rqv,Hidalgo-Duque:2013pva,Guo:2013sya,Baru:2016iwj,Shi:2023mer,Albaladejo:2013aka,Sun:2012zzd,Wang:2020dgr}. 

In $2022$, the Belle collaboration reported a potential isoscalar structure with a mass of $(4014.3\pm 4.0 \pm 1.5)~\mathrm{MeV}$ and a width of $(4\pm 11\pm 6) ~\mathrm{MeV}$ in the $\gamma \psi(2S)$ invariant mass distribution~\cite{Belle:2021nuv}. In view of the proximity to the $D^* \bar{D}^{*}$ threshold, it is a promising candidate for the $D^* \bar{D}^{*}$ bound state. Under the interpretation of the $X_2$ as a $D^*\bar{D}^*$ bound state, the radiative decays $X_2 \to \gamma \psi$ [$\psi = J/\psi, \psi(2S)$] were studied~\cite{Shi:2023mer}. It was found that the ratio of the partial decay width of $X_2 \to \gamma \psi(2S)$ to $X_2 \to \gamma J/\psi$ is smaller than $1.0$, nearly equal to that for the case of $X(3872)$. The mass and width for $X_2$ state predicted in Refs.~\cite{Nieves:2012tt,Guo:2013sya,Albaladejo:2015dsa} match closely with the Belle's measurement~\cite{Belle:2021nuv}.
In Ref.~\cite{Albaladejo:2015dsa}, the hadronic and radiative decays of the $X_2 \to D\bar{D}$, $X_2 \to D\bar{D}^*$, and $X_2 \to D\bar{D}^*\gamma$ were studied using an effective field theory (EFT) approach, and the partial widths of the $X_2 \to D\bar{D}$ and $X_2 \to D\bar{D}^*$ were estimated to be a few MeV and be of the order of keV for $X_2 \to D\bar{D}^*\gamma$. The charmonium decays of the $X_2 \to J/\psi V$ and $X_2 \to \eta_c P$ via the intermediate meson loops, where $V=\rho^0$, $\omega$, and $P=\pi^0$, $\eta$, and $\eta^\prime$ were investigated in Ref.~\cite{Zheng:2024eia}, where the partial decay widths were predicted to be a few tens of keV for $X_2 \to J/\psi \rho^0$, $10^2-10^3$ keV for $X_2 \to J/\psi \omega$, a few keV for $X_2 \to \eta_c \pi^0$, a few tens of keV for $X_2 \to \eta_c \eta$, and a few tenths of keV for $X_2 \to \eta_c \eta^\prime$, respectively.

The theoretical studies mentioned above focus mainly on the charmful decay modes of $X_2$. In order to provide a good platform for better understanding the nature of the $X_2$ state, its charmless decays are also needed. In this work, we investigate the charmless decays of $X_2 \to VV$ and $X_2 \to PP$, where the $X_2$ is assumed to be a pure mesonic molecule of the $D^{*}\bar{D}^{*}$ pair. Using the effective Lagrangian approach, we consider the contributions from the intermediate meson loops. The basic concern of this work is to estimate the partial decay widths of the foregoing processes and to study the affects of the model parameters (such as the phase angle describing the neutral and charged constituent in the $X_2$ and the cutoff in the form factor) and the $X_2$ mass.

The rest of the paper is organized as follows. In Sec.~\ref{sec:formalism}, we present the related decay amplitudes obtained with the effective Lagrangians constructed in the heavy quark limit and chiral symmetry. Then in Sec.~\ref{sec:results} the numerical results and discussions are presented, and a brief summary is given in Sec.~\ref{sec:summary}.

\section{Theoretical Framework} \label{sec:formalism}

\subsection{Effective Lagrangians} \label{subsec:2.1}

We assume that the $X_2$ is an $S$-wave molecular state with the quantum numbers $I(J^{PC})=0(2^{++})$ given by the superposition of $D^{*0}\bar{D}^{*0}$ and $D^{*+}D^{*-}$ hadronic configurations
\begin{equation}\label{eq:WX2}
    |X_2 \rangle = \cos{\theta}|D^{*0}{\bar D}^{*0} \rangle + \sin{\theta}|D^{*+} D^{*-} \rangle,
\end{equation}
where $\theta$ is a phase angle describing the proportion of the neutral and charged constituents. Then, the effective coupling of the $X_2$ state to the $D^*\bar{D}^*$ channel can be written as
\begin{equation}\label{eq:LX2}
    \mathcal{L}_{X_2} = X_{2\mu\nu}\left(\chi_\mathrm{nr}^0  D^{*0 \mu \dagger}\bar{D}^{*0 \nu \dagger} \\+ \chi_{\mathrm{nr}}^\mathrm{c} D^{*+ \mu \dagger}D^{*- \nu \dagger}\right) + \mathrm{H.c.},
\end{equation}
where $\chi_{\mathrm{nr}}^0$ and $\chi_{\mathrm{nr}}^c$ are the coupling constants of the $X_2$ to the neutral and charged $D^*\bar{D}^*$ pairs, respectively. As an isoscalar $D^*\bar{D}^*$ molecular state, the $X_2$ state appears as a pole, $m_{X_2}$, on the real axis in the complex energy plane of the  $T$-matrix obtained from the $D^{*0}\bar{D}^{*0}$ and $D^{*+}D^{*-}$ coupled channels, and the effective couplings $\chi_\mathrm{nr}^0$ and $\chi_\mathrm{nr}^c$ can be derived from the residues of the $T$-matrix elements at the $X_2$ pole~\cite{Meng:2021jnw,Jia:2023pud}:
\begin{eqnarray}
    \chi_{\mathrm{nr}}^0 &=& \left(\frac{16\pi}{\mu^0}\sqrt{\frac{2E_\mathrm{B}^0}{\mu^0}}\right)^{1/2}\cos{\theta}, \label{eq:chi0} \\
    \chi_{\mathrm{nr}}^c &=& \left(\frac{16\pi}{\mu^c}\sqrt{\frac{2E_\mathrm{B}^c}{\mu^c}}\right)^{1/2}\sin{\theta}. \label{eq:chic}
\end{eqnarray}
Here $E_\mathrm{B}^0=m_{D^{*0}}+m_{\bar{D}^{*0}}-m_{X_2}$ and $E_\mathrm{B}^c=m_{D^{*+}}+m_{D^{*-}}-m_{X_2}$ are the binding energies of the $X_2$ relative to the neutral and charged $D^*{\bar D}^*$ threshold, respectively, and $\mu^0=m_{D^{*0}}m_{\bar{D}^{*0}}/(m_{D^{*0}}+m_{\bar{D}^{*0}})$ and $\mu^c=m_{D^{*+}}m_{D^{*-}}/(m_{D^{*+}}+m_{D^{*-}})$ are the reduced masses of $D^{*0}\bar{D}^{*0}$ and $D^{*+}\bar{D}^{*-}$ systems, respectively. Taking the mass of $4.014~\mathrm{GeV}$ of the $X_2$, $\chi_\mathrm{nr}^0$ is $1.32~\mathrm{GeV^{-1/2}} \cos{\theta}$ regarding to the $D^{*0}\bar{D}^{*0}$ component, whereas $\chi_\mathrm{nr}^c$ is $2.36~\mathrm{GeV^{-1/2}} \sin{\theta}$ for the $D^{*+}D^{*-}$ component. The different couplings due to the different masses between the charged and neutral charmed mesons would lead to an isospin-breaking effect. 

Based on the heavy quark limit and chiral symmetry, the effective Lagrangian involving the light vector and pseudoscalar mesons can be constructed as~\cite{Wu:2021udi,Wang:2022qxe,Wu:2022hck,Zheng:2024eia}
\begin{align}\label{eq:LDDVP}
\mathcal{L} &= -\ii g_{DDV} D_i^\dagger \pararrowk{\mu}D^{j}(V_\mu^\dagger)_j^i \nonumber\\
&-2 f_{D^*DV}\epsilon_{\mu\nu\alpha\beta} (\partial^\mu V^{\nu\dagger})_j^i(D_i^\dagger \pararrowk{\alpha}D^{*\beta j} \nonumber\\
&- D_i^{*\beta\dagger}\pararrowk{\alpha}D^{j})+\ii g_{D^*D^*V}D_i^{*\nu\dagger}\pararrowk{\mu}D_\nu^{*j}(V_\mu^\dagger)_j^i\nonumber \\
&+ \ii 4 f_{D^*D^*V} D_{i\mu}^{*\dagger}(\partial^\mu V^{\nu\dagger} - \partial^\nu V^{\mu\dagger})_j^i D_\nu^{*j}\nonumber\\
&-\ii g_{D^*DP}\big(D^{i \dagger}\partial^{\mu} P_{ij}^\dagger D_\mu^{*j} - D_\mu^{*i\dagger}\partial^\mu P_{ij}^\dagger D^j\big) \nonumber\\
&+ \frac{1}{2} g_{D^*D^*P}\epsilon_{\mu\nu\alpha\beta} D_i^{*\mu\dagger}\partial^\nu P^{ij\dagger}\pararrowk{\alpha} D_j^{*\beta},
\end{align}
where $D^{(*)} = (D^{(*)0}, D^{(*)+}, D_s^{(*)+})$ and $D^{(*)\dagger} = (\bar{D}^{(*)0}, D^{(*)-}, D_s^{(*)-})$. The $V$ and $P$ are, respectively, the nonet vector and pseudoscalar mesons in the following matrix form:
\begin{subequations}\label{eq:vmatrix}
\begin{align}
    V &= 
    \begin{pmatrix}
    \frac{\rho^0}{\sqrt{2}}+\frac{\omega}{\sqrt{2}}&\rho^+&K^{*+}\\
    \rho^-&-\frac{\rho^0}{\sqrt{2}}+\frac{\omega}{\sqrt{2}}&K^{*0}\\
    K^{*-}&\bar{K}^{*0}&\phi
    \end{pmatrix},\\
	P &= \begin{pmatrix}
		\frac{\pi^0}{\sqrt{2}} + \frac{\delta \eta + \gamma \eta'}{\sqrt{2}} & \pi^+ & K^+\\
		\pi^- & -\frac{\pi^0}{\sqrt{2}} + \frac{\delta \eta + \gamma \eta'}{\sqrt{2}} & K^0\\
		K^- & \bar{K}_0 & - \gamma \eta + \delta \eta' \label{eq:P}
	\end{pmatrix}.
\end{align}
\end{subequations}
Here $\delta=\cos(\theta_\mathrm{P}+\arctan\sqrt{2})$ and $\gamma=\sin(\theta_\mathrm{P}+\arctan\sqrt{2})$ with the $\eta$-$\eta'$ mixing angle $\theta_\mathrm{P}$ ranging from $-24.6^\circ$ to $-11.5^\circ$ \cite{2022pdgPoTaEP2022-083C01}.

The coupling constants of the charmed meson to the light vector and pesudoscalar mesons have the following relationship~\cite{Casalbuoni:1996pg,Cheng:2004ru}
\begin{subequations}\label{eq:gddvs}
\begin{align}
    g_{DDV} &= g_{D^*D^*V} = \frac{\beta g_V}{\sqrt{2}},\label{eq:gddvgdsdsv}\\
    f_{D^*DV}&= \frac{f_{D^*D^*V}}{m_{D^*}} = \frac{\lambda g_V}{\sqrt{2}},\label{eq:fdsdvfdsdsv}\\
    g_{D^*D^*P} &= \frac{g_{D^*DP}}{\sqrt{m_Dm_{D^*}}}= \frac{2g}{f_{\pi}},\label{eq:gddp}
\end{align}
\end{subequations}
where $\beta=0.9$, $\lambda=0.56~\mathrm{GeV^{-1}}$, $g = 0.59$~\cite{Isola:2003fh}, and $g_V = m_\rho / f_\pi$ with the pion decay constant $f_\pi = 132~\mathrm{MeV}$~\cite{Casalbuoni:1996pg} and the $\rho$ meson mass $m_{\rho}=775.26~\mathrm{MeV}$~\cite{2022pdgPoTaEP2022-083C01}.

\subsection{Transition amplitudes of $X_2\to V V$ and $X_2\to P P$} \label{subsec:2.2}

\begin{figure}[htbp]
    \includegraphics[width=0.85\linewidth]{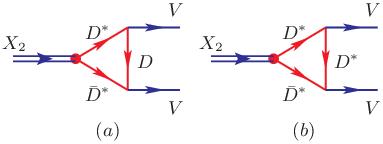}\,  \\
	\includegraphics[width=0.85\linewidth]{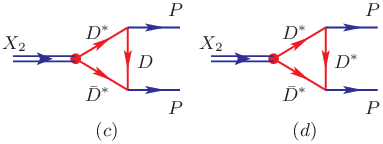}
    \caption{Feynman diagrams for the processes $X_2 \to VV$[(a)-(b)] and $X_2 \to PP$[(c)-(d)] via charmed meson loops.}
    \label{fig:Feynmandiagrams}
\end{figure}

According to the effective Lagrangians above, the decays of $X_2 \to VV$ and $X_2 \to PP$ can occur via the charmed meson loops as shown in Fig.~\ref{fig:Feynmandiagrams}. The decay amplitudes for $X_2(p)\to [D^*(p_1){\bar D}^*(p_2)]D^{(*)}(q)\to V_1(p_3)V_2(p_4)$ and $X_2(p)\to [D^*(p_1){\bar D}^*(p_2)]D^{(*)}(q)\to P_1(p_3)P_2(p_4)$ (the particle outside the parentheses is the exchanged particle, and $p$, $p_1$, $p_2$, $q$, $p_3$, and $p_4$ are the four-momentum of corresponding particles) can be written as
	\begin{eqnarray}
		\mathcal{M}_V \!\! &=& \!\! \chi_{\mathrm{nr}}^{0,c }\sqrt{m_{X_2}}m_{D^*} \varepsilon^{\mu\nu}(X_2)\varepsilon^{*\alpha}(V_1)\varepsilon^{*\beta}(V_2) I_{\mu\nu\alpha\beta},\\
		\mathcal{M}_P \!\! &=& \!\! \chi_{\mathrm{nr}}^{0,c }\sqrt{m_{X_2}}m_{D^*}\varepsilon^{\mu\nu}(X_2) I_{\mu\nu}.
	\end{eqnarray}
The factor $\sqrt{m_{X_2}}m_{D^*}$ accounts for the nonrelativistic normalization of the heavy fields involved in the $X_2D^*\bar{D}^*$ vertex. The $\varepsilon^{\mu\nu}(X_2)$, $\varepsilon^{*\alpha}(V_1)$, and $\varepsilon^{*\beta}(V_2) $ describe the polarization tensor of the initial state $X_2$, the polarization vectors of the final state $V_1$, and $V_2$, respectively. The tensor structures $I_{\mu\nu\alpha\beta}$ and $I_{\mu\nu}$ are expressed as
    \begin{align}
	I_{\mu\nu\alpha\beta}^{a} &= \int \frac{\mathrm{d}^4q}{(2\pi)^4} g_{\mu\rho}g_{\nu\sigma} [-2f_{D^*DV}\epsilon_{\delta\alpha\omega\xi}p_3^\delta(p_1+q)^\omega]\nonumber\\
	&\times [2f_{D^*DV}\epsilon_{\lambda\beta\gamma\eta}p_4^\lambda(q-p_2)^\gamma]S^{\rho\xi}(p_1,m_{D^*})\nonumber\\
	&\times S^{\sigma\eta}(p_2,m_{D^*}) S(q,m_D)F(q^2,m_D^2),\\
	I_{\mu\nu\alpha\beta}^{b} &= \int \frac{\mathrm{d}^4q}{(2\pi)^4} g_{\mu\rho}g_{\nu\sigma} [4f_{D^*D^*V}(p_{3\delta} g_{\alpha\xi} - p_{3\xi} g_{\alpha\delta})\nonumber\\
	&- g_{D^*D^*V}(p_1+q)_\alpha g_{\xi\delta}] [4f_{D^*D^*V}(p_{4\eta} g_{\beta\gamma} - p_{4\gamma} g_{\beta\eta}) \nonumber\\
	&+ g_{D^*D^*V}(p_2-q)_\beta g_{\eta\gamma}]S^{\rho\xi}(p_1,m_{D^*}) \nonumber\\
	&\times S^{\sigma\eta}(p_2,m_{D^*}) S^{\delta\gamma}(q,m_{D^*})F(q^2,m_{D^*}^2),\\
	I_{\mu\nu}^{c}&= \int \frac{\mathrm{d}^4q}{(2\pi)^4} g_{\mu\rho}g_{\nu\sigma} [-g_{D^*DP} p_{3\xi}] \nonumber\\
	&\times [g_{D^*DP} p_{4\eta}] S^{\rho\xi}(p_1,m_{D^*}) \nonumber\\
	&\times S^{\sigma\eta}(p_2,m_{D^*})S(q,m_D)F(q^2,m_D^2),\\
	I_{\mu\nu}^{d}&= \int \frac{\mathrm{d}^4q}{(2\pi)^4} g_{\mu\rho}g_{\nu\sigma} [\frac{1}{2} g_{D^*D^*P} \epsilon_{\delta\lambda\omega\xi}p_3^\lambda (p_1+q)^\omega]\nonumber\\
	&\times [\frac{1}{2}g_{D^*D^* P} \epsilon_{\eta\theta\kappa\gamma}p_4^\theta(q-p_2)^\kappa]S^{\rho\xi}(p_1,m_{D^*}) \nonumber\\
	&\times S^{\sigma\eta}(p_2,m_{D^*}) S^{\delta\gamma}(q,m_{D^*})F(q^2,m_{D^*}^2).
    \end{align}
Here $S(q,m_D)$ and $S^{\mu\nu}(q,m_{D^*})$ represent the propagators for the charmed mesons $D$ and $D^*$ with the following forms, respectively,
\begin{eqnarray}
		S(q,m_D) &=& \frac{1}{q^2-m_D^2+\ii \epsilon},\\
		S^{\mu\nu}(q,m_{D^*}) &=& \frac{-g^{\mu\nu} + q^\mu q^\nu/m_{D^*}^2}{q^2-m^2_{D^*}+\ii \epsilon}.
\end{eqnarray}

The $F(q^2,m^2)$ is a form factor to consider the off-shell effect and the inner structure of the exchanged particle~\cite{Gortchakov:1995im,Tornqvist:1993ng,Cheng:2004ru,Li:1996yn,Locher:1993cc}. Because the mass of the $X_2$ is close to the thresholds of the $D^{*}\bar{D}^{*}$, the two charmed mesons $D^*$ and $\bar{D}^*$ interacting with the $X_2$ could be considered to be nearly on-shell. However, the exchanged charmed meson $D$ or $D^*$ in the triangle loop is off-shell. In this work we adopt a dipole form factor
\begin{equation}
    F(q^2,m^2) = \left(\frac{m^2-\Lambda^2}{q^2-\Lambda^2}\right)^2,
\end{equation}
which is normalized to unity at $q^2=m^2$~\cite{Cheng:2004ru}, where $q$ and $m$ are the momentum and mass of the exchanged mesons, and the cutoff $\Lambda$ can be further reparameterized as $\Lambda = m + \alpha \Lambda_{\mathrm{QCD}}$ with $\Lambda_{\mathrm{QCD}} = 0.22~\mathrm{GeV}$~\cite{Cheng:2004ru}, in which the model parameter $\alpha$ is usually expected to be of the order of unity~\cite{Cheng:2004ru,Tornqvist:1993vu,Tornqvist:1993ng,Locher:1993cc,Li:1996yn}. In the present calculations, we take $\alpha$ ranging from $0.6$ to $1.2$.

With all the ingredients above, the partial decay width for the $X_2 \to VV (PP)$ decay is given by
\begin{align}
\Gamma= \frac{1}{5S}\frac{\vert \vec{p}_3 \vert}{8 \pi m_{X_2}^2} \sum_{\rm spins} \left\vert \mathcal{M}_{V(P)}\right\vert^2,
\label{Eq.X2 decay rate}
\end{align}
where the symmetry factor $S$ is taken to be 2 for the decays having the identical particles in the final states, and to be 1 for other cases. The symbol $\sum_{\rm spins}$ means the summation over the polarizations of the initial $X_2$ and final vector mesons.

\section{Numerical Results And Discussions}\label{sec:results}

In the following, we present the partial decay widths of the $X_2 \to VV$ and $X_2 \to PP$. We select three different phase angles $\theta = 0$, $\pi$/6, and $\pi$/4. With the phase angle $\theta = 0$, the $X_2$ contains only neutral component. For $\theta = \pi/6$, the neutral component is dominant in the $X_2$, while for $\theta = \pi/4$, the proportions of neutral and charged constituents are equal. Besides, we take the $\eta$-$\eta'$ mixing angle $\theta_\mathrm{P}$ to be $-19.1^\circ$~\cite{MARK-III:1988crp,DM2:1988bfq}.

\begin{figure*}[htbp]
	\includegraphics[width=0.32\linewidth]{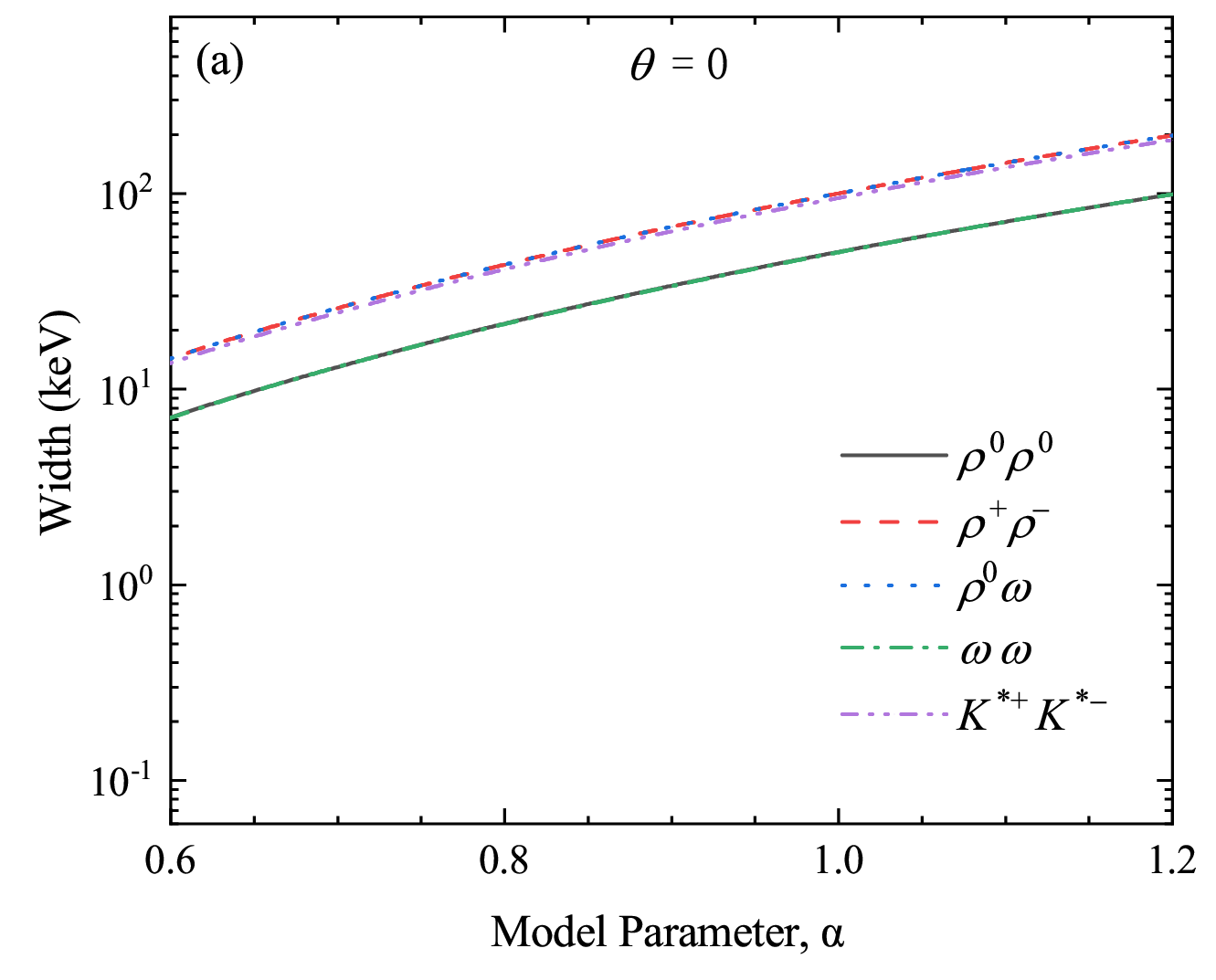}\, 
	\includegraphics[width=0.32\linewidth]{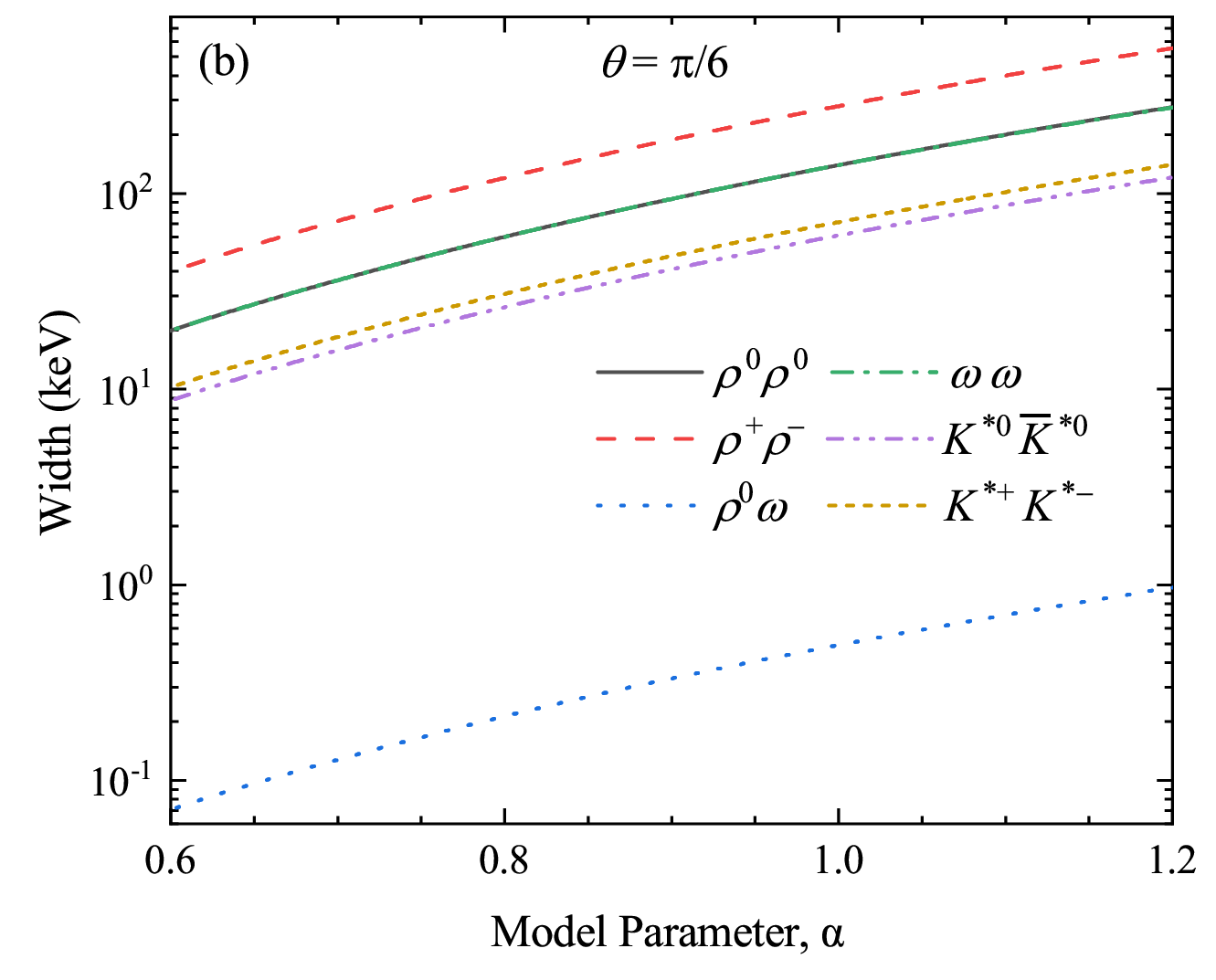}\, 
	\includegraphics[width=0.32\linewidth]{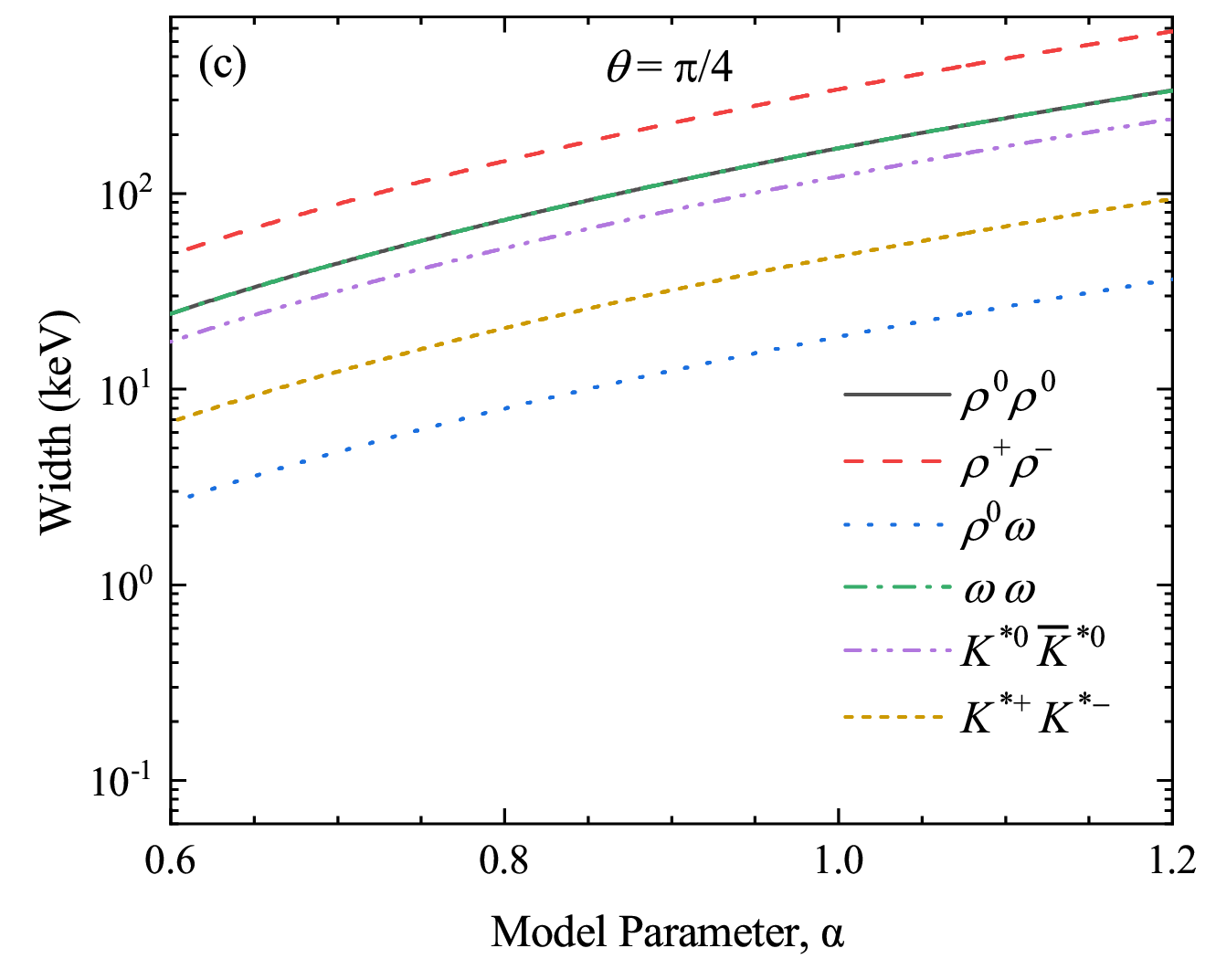}
	\caption{The decay widths of the $X_2 \to V V$ as a function of the model parameter $\alpha$. As indicated, three different phase angles $\theta = 0, \pi/6$, and $\pi/4$ are chosen.}
	\label{fig:VVθ03045alpha}
\end{figure*}

In Fig.~\ref{fig:VVθ03045alpha}, we show the $\alpha$ dependence of the partial decay widths of $X_2 \to V V$ for different phase angles. The $X_2$ decays to $K^{*+} K^{*-}$ via the $[D^{*0} \bar{D}^{*0}]D_s^{(*)+}$ intermediate mesons, while it decays to $K^{*0} \bar{K}^{*0}$ via the $[D^{*+} D^{*-}]D_s^{(*)+}$ intermediate mesons. As a result, in the case of $\theta = 0$ for which the $X_2$ is the state made of only the neutral charmed $D^{*0}\bar{D}^{*0}$, there is no neutral $K^{*0} \bar{K}^{*0}$ channel [see Fig. \ref{fig:VVθ03045alpha}(a)]. It is clearly seen that all the widths increase with increasing the model parameter $\alpha$. In the range of $\alpha = 0.6-1.2$, the predicted partial decay widths of $X_2 \to V V$ can reach several hundred keV.

With the phase angle $\theta = 0$, the $X_2$ has only neutral charmed component so that the isospin-violating decay $X_2 \to \rho^0 \omega$ occurs with almost the same rate as that of the $X_2 \to \rho^+ \rho^-$ and $X_2 \to K^{*+} K^{*-}$. The partial decay width of $X_2\to \rho^0 \rho^0$ almost equals to that of $X_2 \to \omega \omega$, and it is about two times smaller than that of $X_2 \to \rho^0 \omega$. 

In the case of $\theta = \pi/6$, the effective couplings $\chi_\mathrm{nr}^0\approx\chi_\mathrm{nr}^c$ according to Eqs. \eqref{eq:chi0} and \eqref{eq:chic}. As a consequence, the contributions from the charged and neutral charmed meson loops are nearly equal. The isospin-violated process of the $X_2$ decaying into $\rho^0 \omega$ is highly suppressed. As seen in Fig.~\ref{fig:VVθ03045alpha}(b), the partial width of $X_2 \to \rho^0 \omega$ is about two orders of magnitude smaller than those of other decay modes.

When the phase angle $\theta$ increases towards to $\pi/4$, the contributions from the charged charmed meson loops would become more important than the neutral ones, thereby increasing the decay rates of the processes $X_2\to\rho^0\omega$ and $X_2\to K^{*0}\bar{K}^{*0}$, but reducing the rate of the $X_2\to K^{*+}K^{*-}$. Eventually, the partial decay width of the $X_2\to K^{*0}\bar{K}^{*0}$ becomes larger than that of the $X_2 \to K^{*+}K^{*-}$, exhibiting contrary behavior to the case of $\theta = \pi/6$. It is noted that the partial decay width of $X_2 \to \rho^0 \rho^0(\omega \omega)$ and $X_2 \to \rho^+ \rho^-$ are not sensitive to the phase angle $\theta$ except $\theta = 0$.

\begin{figure*}[htbp]
	\includegraphics[width=0.32\linewidth]{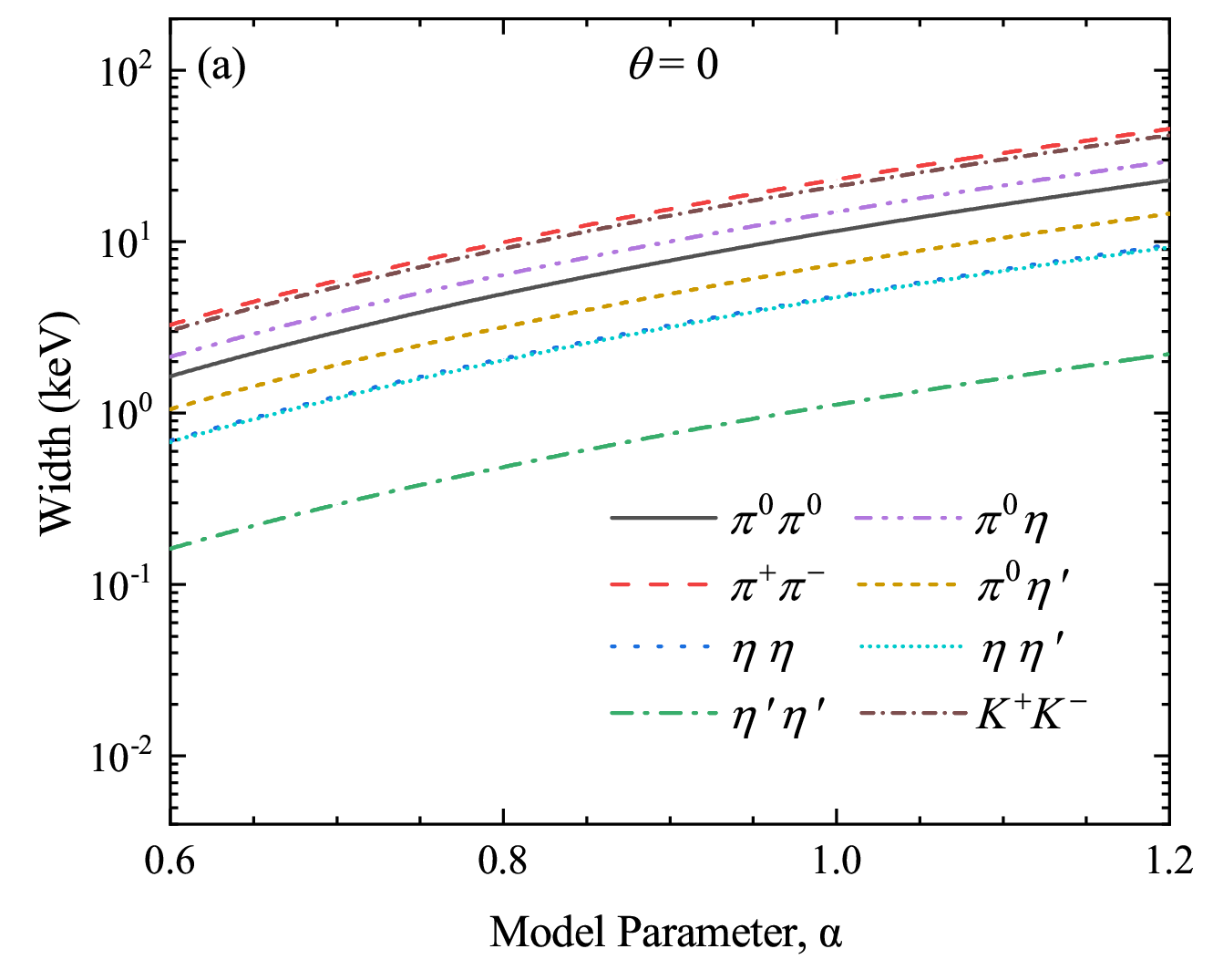}\, 
	\includegraphics[width=0.32\linewidth]{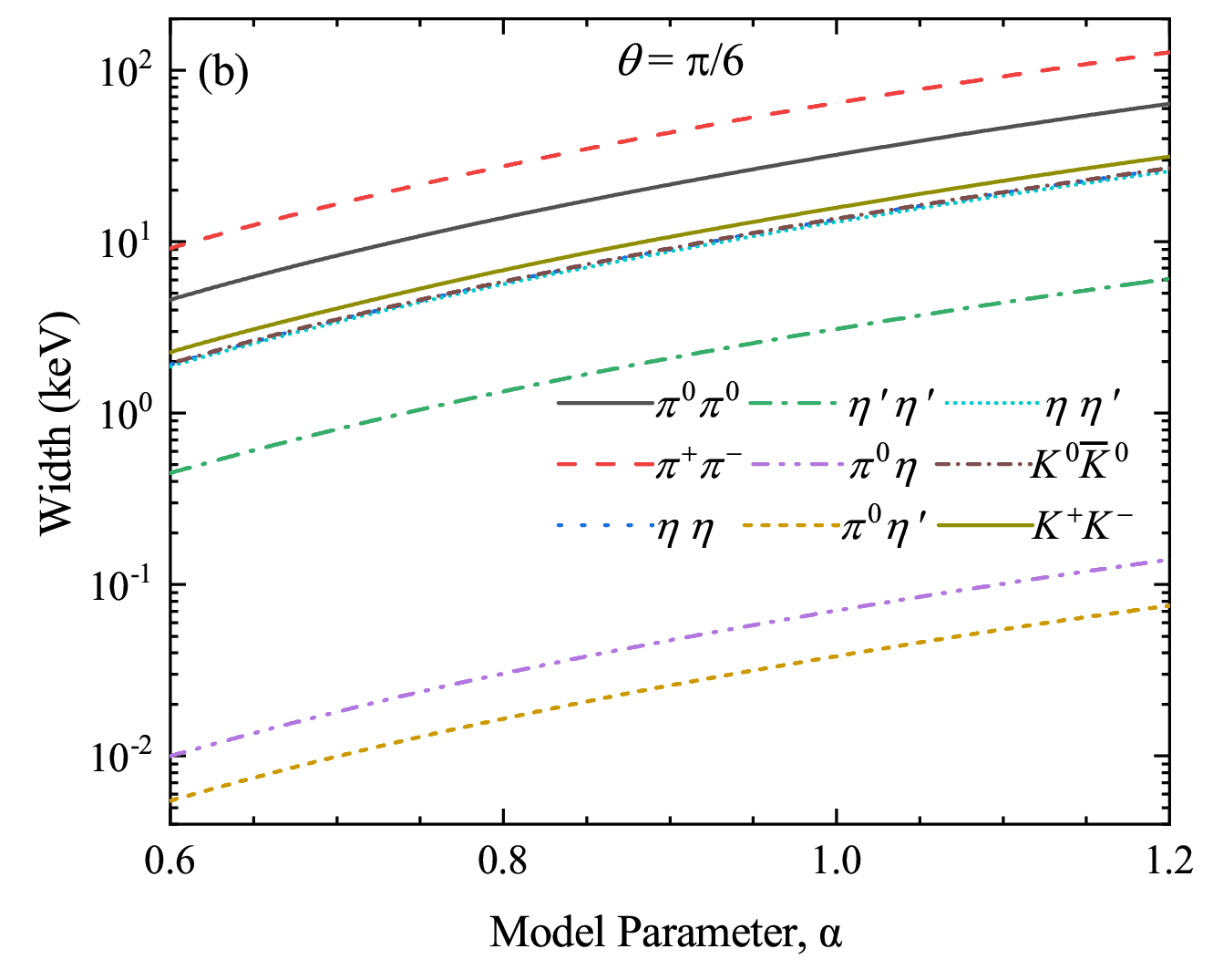}\,
	\includegraphics[width=0.32\linewidth]{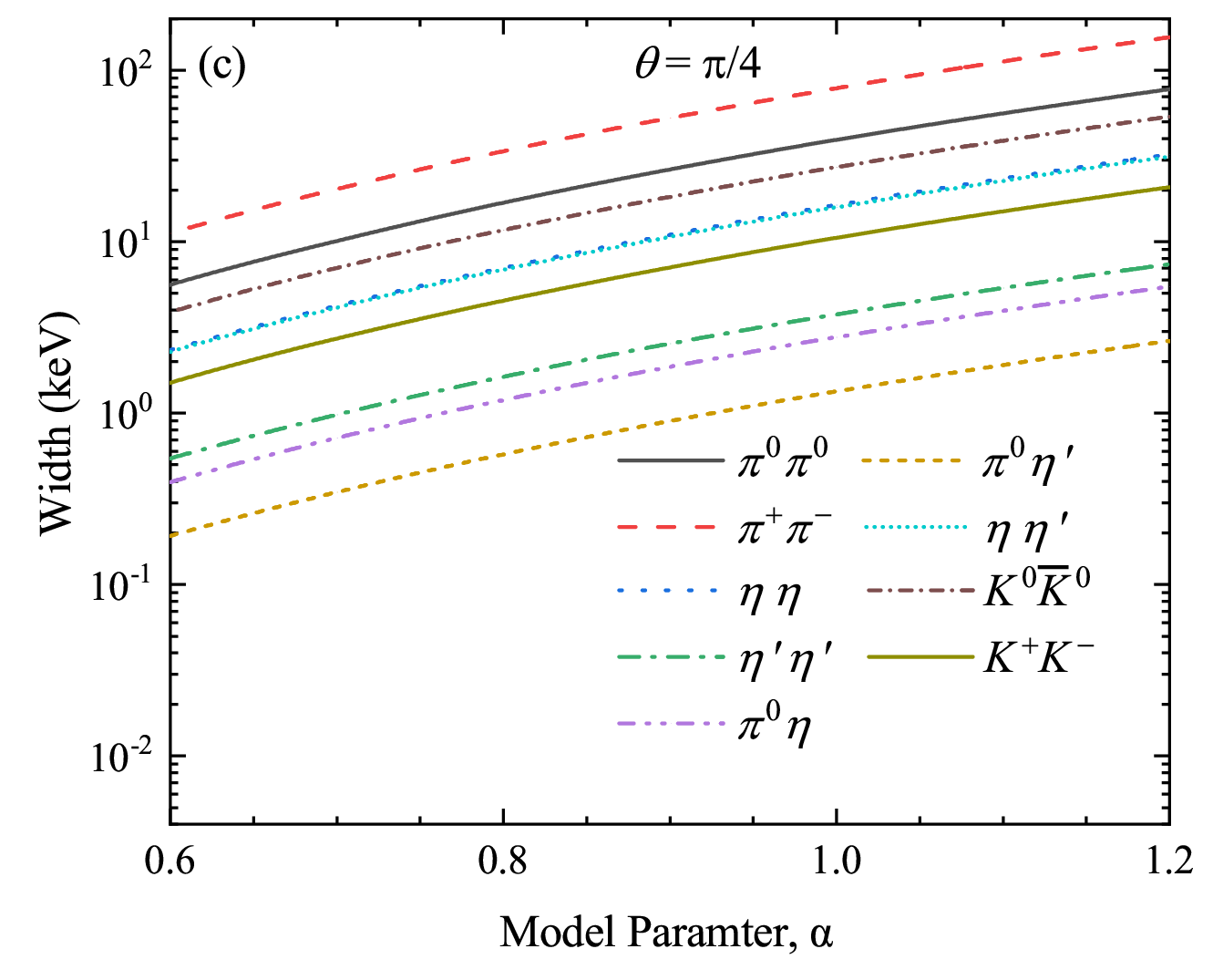}
	\caption{The decay widths of the $X_2 \to P P$ as a function of the model parameter $\alpha$. As indicated, three different phase angles $\theta = 0, \pi/6$, and $\pi/4$ are chosen. The $\eta$-$\eta^\prime$ mixing angle is taken as $\theta_\mathrm{P} = -19.1^\circ$.}
	\label{fig:PPθ03045alpha}
\end{figure*} 

In Fig.~\ref{fig:PPθ03045alpha}, we plot the obtained partial decay widths of $X_2 \to P P$ as a function of the model parameter $\alpha$. It is seen that the behavior of the partial widths for the processes $X_2\to PP$ with varying the the model parameter $\alpha$ and the phase angle $\theta$ are similar to those for the $X_2\to VV$. On the whole, the $X_2$ goes to the $VV$ at a higher rate than to the $PP$. According to the theoretical results of the partial decay widths shown in Figs.~\ref{fig:VVθ03045alpha} and \ref{fig:PPθ03045alpha}, we summarize in Table \ref{tab:branchingratio} the branching fractions for the $X_2\to VV$ and $X_2\to PP$ using the measured total width $\Gamma_\mathrm{t}=4~\mathrm{MeV}$ by the Belle collaboration \cite{Belle:2021nuv}.

\begin{table*}[htbp]
	\caption{The branching ratios for $X_2 \to V V$ and $X_2 \to P P$ with different $\theta$ values. Here the $\alpha$ range is taken to be $0.6–1.2$.}
	\label{tab:branchingratio}
	\begin{ruledtabular}
		\begin{tabular}{cccc}
		Final states	&$\theta = 0$& $\theta = \pi/6$&$\theta = \pi/4$\\
			\colrule
			$\rho^0 \rho^0$ & $(0.18-2.48)\times 10^{-2}$ & $(0.50-6.89)\times 10^{-2}$ & $(0.61-8.40)\times 10^{-2}$\\
			$\rho^+ \rho^-$ & $(0.36-4.95)\times 10^{-2}$ & $(0.10-1.38)\times 10^{-1}$ & $(0.12-1.68)\times 10^{-1}$\\
			$\rho^0 \omega$ & $(0.36-4.95)\times 10^{-2}$ & $(0.18-2.42)\times 10^{-4}$ & $(0.66-9.11)\times 10^{-3}$\\
			$\omega \omega$ & $(0.18-2.47)\times 10^{-2}$ & $(0.50-6.88)\times 10^{-2}$ & $(0.61-8.38)\times 10^{-2}$\\
			$K^{*0} \bar{K}^{*0}$ & $\cdots$ & $(0.22-3.01)\times 10^{-2}$ & $(0.44-6.02)\times 10^{-2}$\\
			$K^{*+} K^{*-}$ & $(0.34-4.68)\times 10^{-2}$ & $(0.26-3.51)\times 10^{-2}$ & $(0.17-2.34)\times 10^{-2}$\\
			$\pi^0 \pi^0$ & $(0.41-5.69)\times 10^{-3}$ & $(0.11-1.59)\times 10^{-2}$ & $(0.14-1.94)\times 10^{-2}$\\
			$\pi^+ \pi^-$ & $(0.08-1.14)\times 10^{-2}$ & $(0.23-3.18)\times 10^{-2}$ & $(0.28-3.88)\times 10^{-2}$\\
			$\eta \eta$ & $(0.17-2.38)\times 10^{-3}$ & $(0.48-6.62)\times 10^{-3}$ & $(0.58-8.06)\times 10^{-3}$\\
			$\eta' \eta'$ & $(0.40-5.51)\times 10^{-4}$ & $(0.11-1.52)\times 10^{-3}$ & $(0.14-1.85)\times 10^{-3}$\\
			$\pi^0 \eta$ & $(0.53-7.37)\times 10^{-3}$ & $(0.25-3.48)\times 10^{-5}$ & $(0.10-1.37)\times 10^{-3}$\\
			$\pi^0 \eta'$ & $(0.26-3.63)\times 10^{-3}$ & $(0.14-1.88)\times 10^{-5}$ & $(0.48-6.58)\times 10^{-4}$\\
			$\eta \eta'$ & $(0.17-2.32)\times 10^{-3}$ & $(0.47-6.42)\times 10^{-3}$ & $(0.57-7.82)\times 10^{-3}$\\
			$K^{0} \bar{K}^{0}$ & $\cdots$ & $(0.49-6.72)\times 10^{-3}$ & $(0.10-1.34)\times 10^{-2}$\\
			$K^{+} K^{-}$ & $(0.08-1.04)\times 10^{-2}$ & $(0.56-7.80)\times 10^{-3}$ & $(0.38-5.20)\times 10^{-3}$\\
		\end{tabular}
	\end{ruledtabular}
\end{table*}

\begin{figure*}[htbp]
	\centering
	\includegraphics[width=0.32\linewidth]{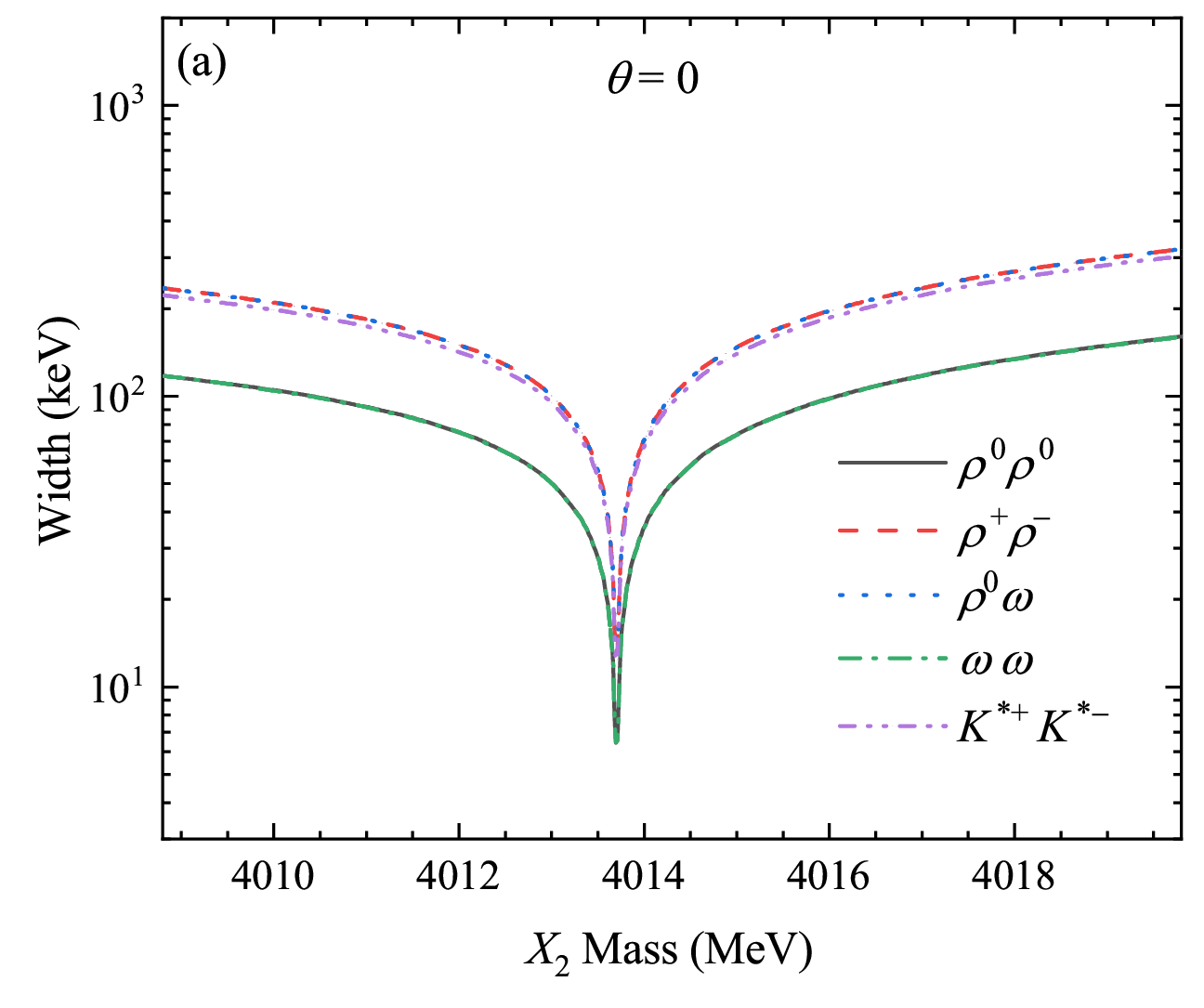}\,
	\includegraphics[width=0.322\linewidth]{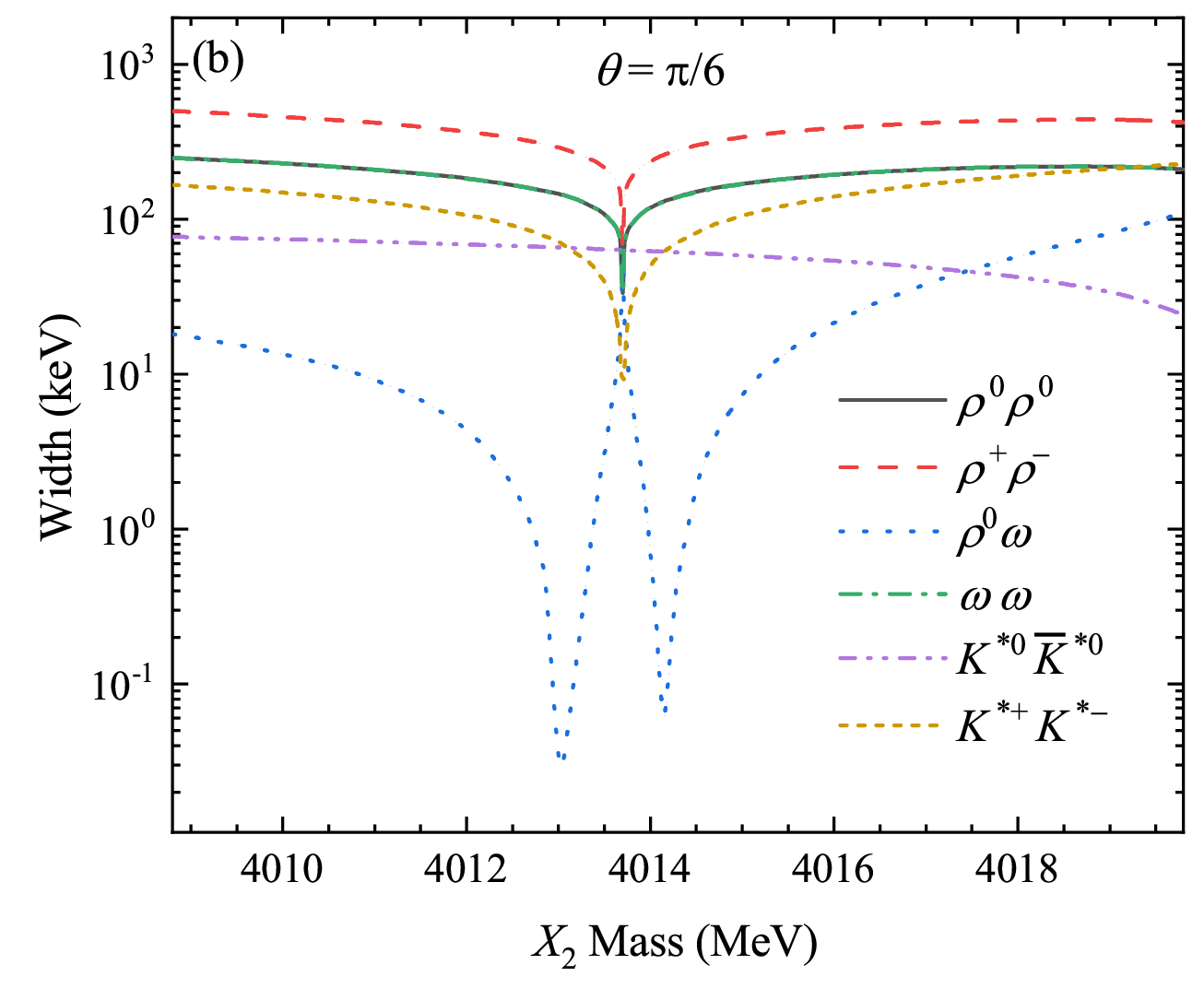}\,
	\includegraphics[width=0.32\linewidth]{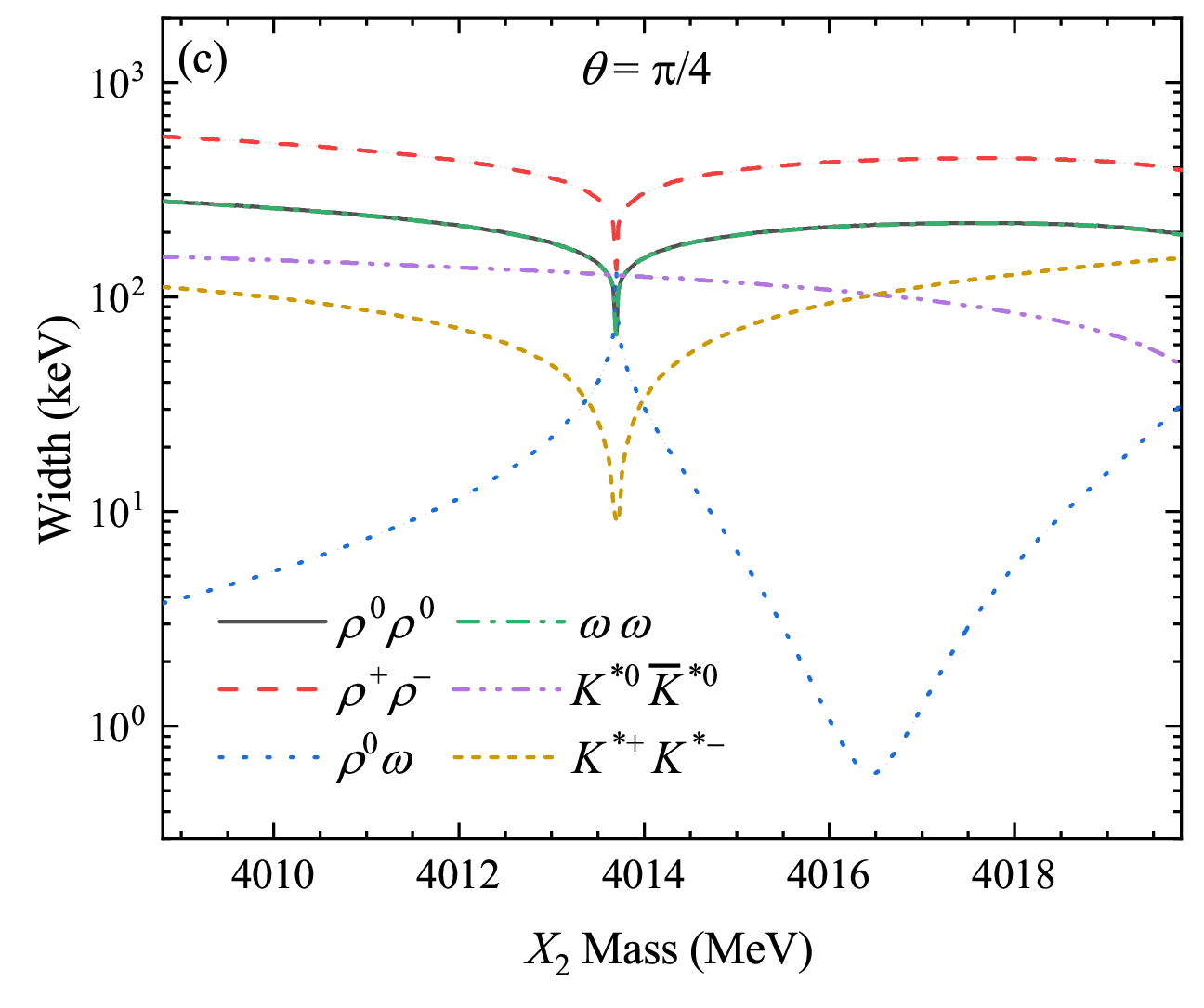}\,
	\caption{The $X_2$ mass dependence of the decay processes $X_2 \to V V$ for different phase angles. The model parameter $\alpha$ is taken to be 1.0.}
	\label{fig:VVm}
\end{figure*}

\begin{figure*}
	\centering
	\includegraphics[width=0.32\linewidth]{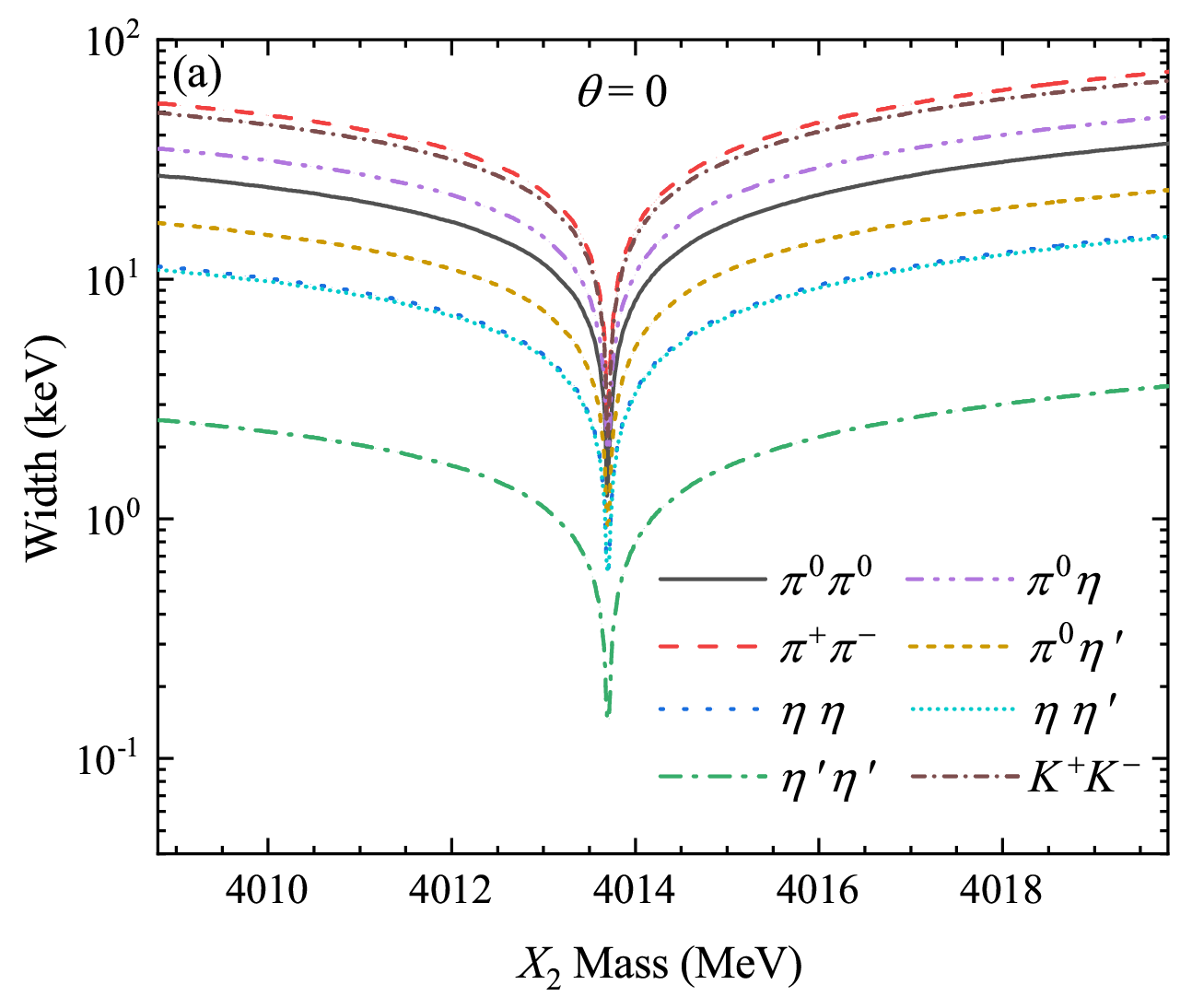}\,
	\includegraphics[width=0.32\linewidth]{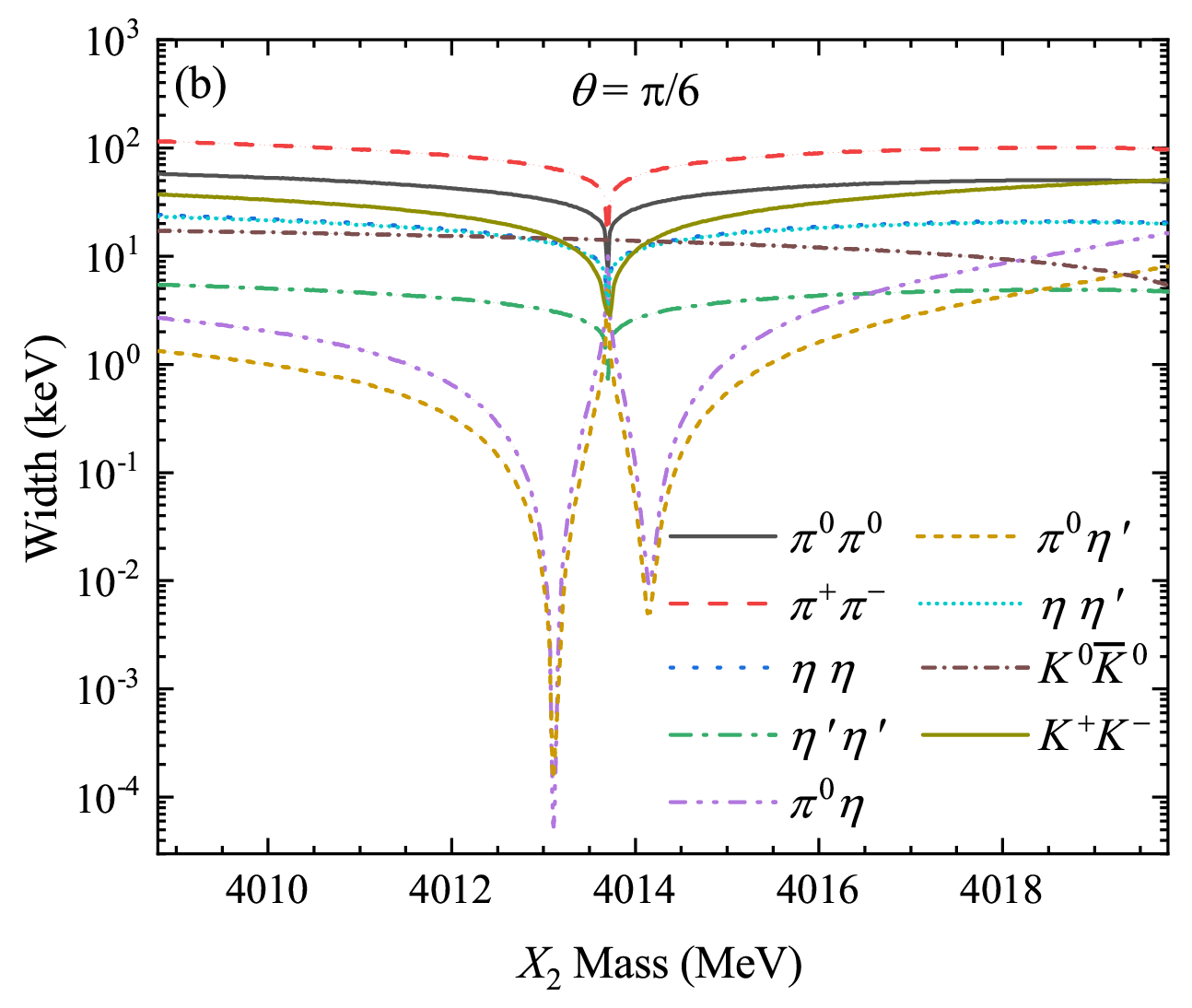}\,
	\includegraphics[width=0.32\linewidth]{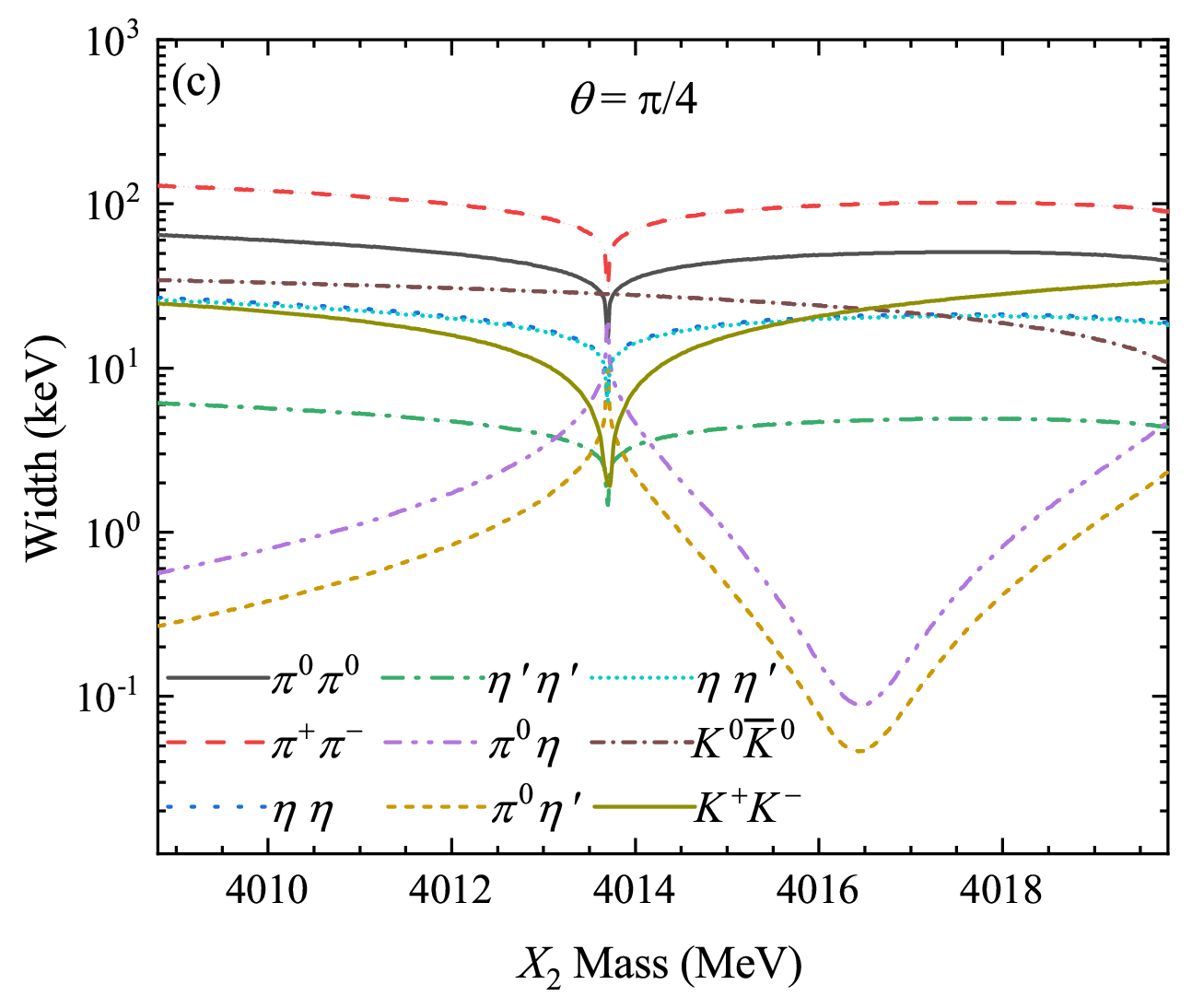}\,
	\caption{The $X_2$ mass dependence of the decay processes $X_2 \to P P$ for different phase angles. The model parameter $\alpha$ is taken to be 1.0 and the $\eta$-$\eta^\prime$ mixing angle $\theta_\mathrm{P} = -19.1^\circ$.}
	\label{fig:PPm}
\end{figure*}

To study the $X_2$ mass dependence of the decay processes we consider, we vary the $X_2$ mass from $4.009~\mathrm{GeV}$ to $4.020~\mathrm{GeV}$ in view of the mass value $(4014.3\pm 4.0 \pm 1.5)~\mathrm{MeV}$ measured by the Belle collaboration \cite{Belle:2021nuv}. The calculated the partial decay widths for different $X_2$ masses are shown in Fig.~\ref{fig:VVm} and Fig.~\ref{fig:PPm}. In the calculations, we fixed the model parameter $\alpha=1.0$ and again choose three phase angles $\theta=0,\,\pi/6,\,\pi/4$. It is noted that the partial decay widths for the $X_2\to VV$ and $X_2\to PP$ exhibit similar behavior with varying the $X_2$ mass.

Near the $D^{*0} \bar{D}^{*0}$ threshold, namely $m_{X_2} = 4.0137~\mathrm{GeV}$, the effective coupling $\chi_\mathrm{nr}^0$ in Eq. \eqref{eq:chi0} approaches zero so that the contribution from the neutral charmed meson loops is negligible. Hence, the partial widths for the isospin violated processes, which are governed by the difference between the neutral and charged meson loops, show a peak near the $D^{*0} \bar{D}^{*0}$ threshold, while the isospin conserved ones exhibit a valley [see Figs.~\ref{fig:VVm} (b) and (c) and Fig.~\ref{fig:PPm} (b) and (c)]. However, in the special case of $\theta=0$, the $X_2$ has no charged component, thus the isospin violated and conserved processes vary with the $X_2$ mass in the similar manner. The valleys for the isospin violated processes near $m_{X_2} = 4.0128~\mathrm{GeV}$ and $m_{X_2} = 4.0144~\mathrm{GeV}$ in Figs. ~\ref{fig:VVm}(b) and \ref{fig:PPm}(b), and near $m_{X_2} = 4.0171~\mathrm{GeV}$ in Figs. ~\ref{fig:VVm}(c) and \ref{fig:PPm}(c) could be reproduced by Eq. (13) in Ref.~\cite{Zheng:2024eia}. Our calculations indicate that the partial decay widths for the isospin conserved processes are not very sensitive to the $X_2$ masses, unless the mass of $X_2$ is very closed to the $D^{*}\bar{D}^{*}$ mass threshold.

\begin{figure}
	\centering
	\includegraphics[width=1.0\linewidth]{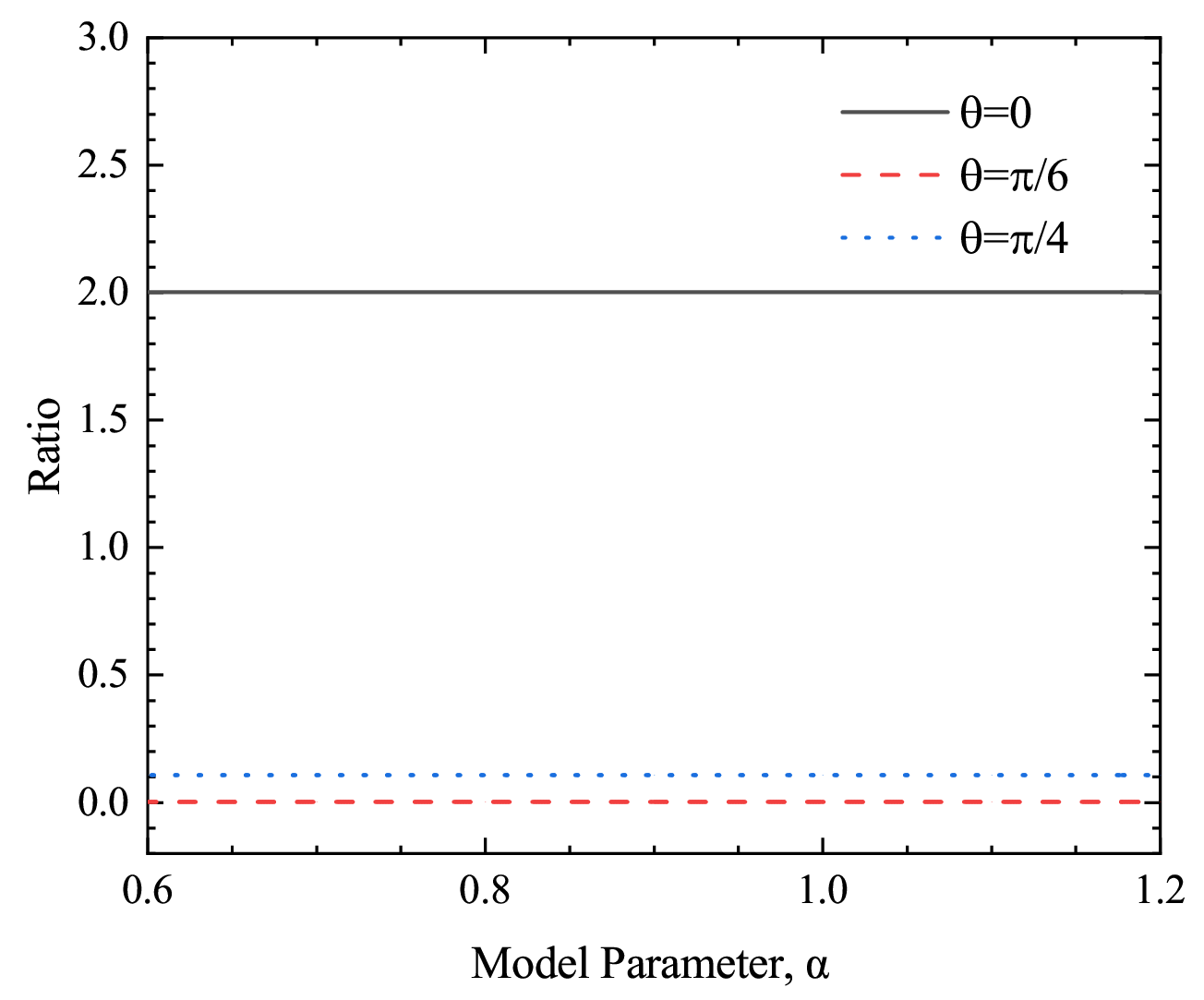}
	\caption{The $\alpha$ dependence of the ratio $R_1$ defined in Eq.~(\ref{eq:ratio_VV1}).}
	\label{fig:VV_PPalpha}
\end{figure}

\begin{figure*}
	\centering
	\includegraphics[width=0.45\linewidth]{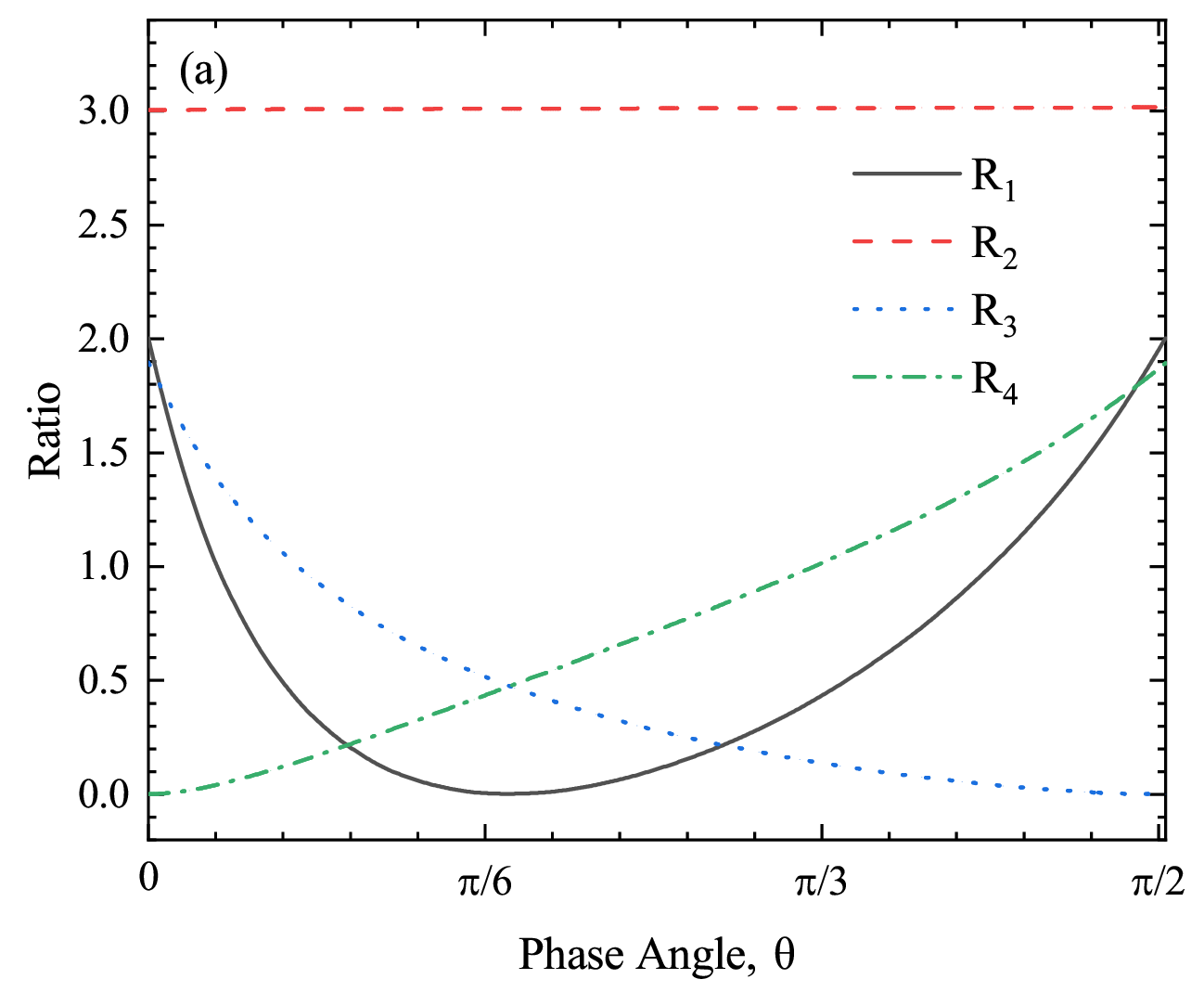}\,
	\includegraphics[width=0.45\linewidth]{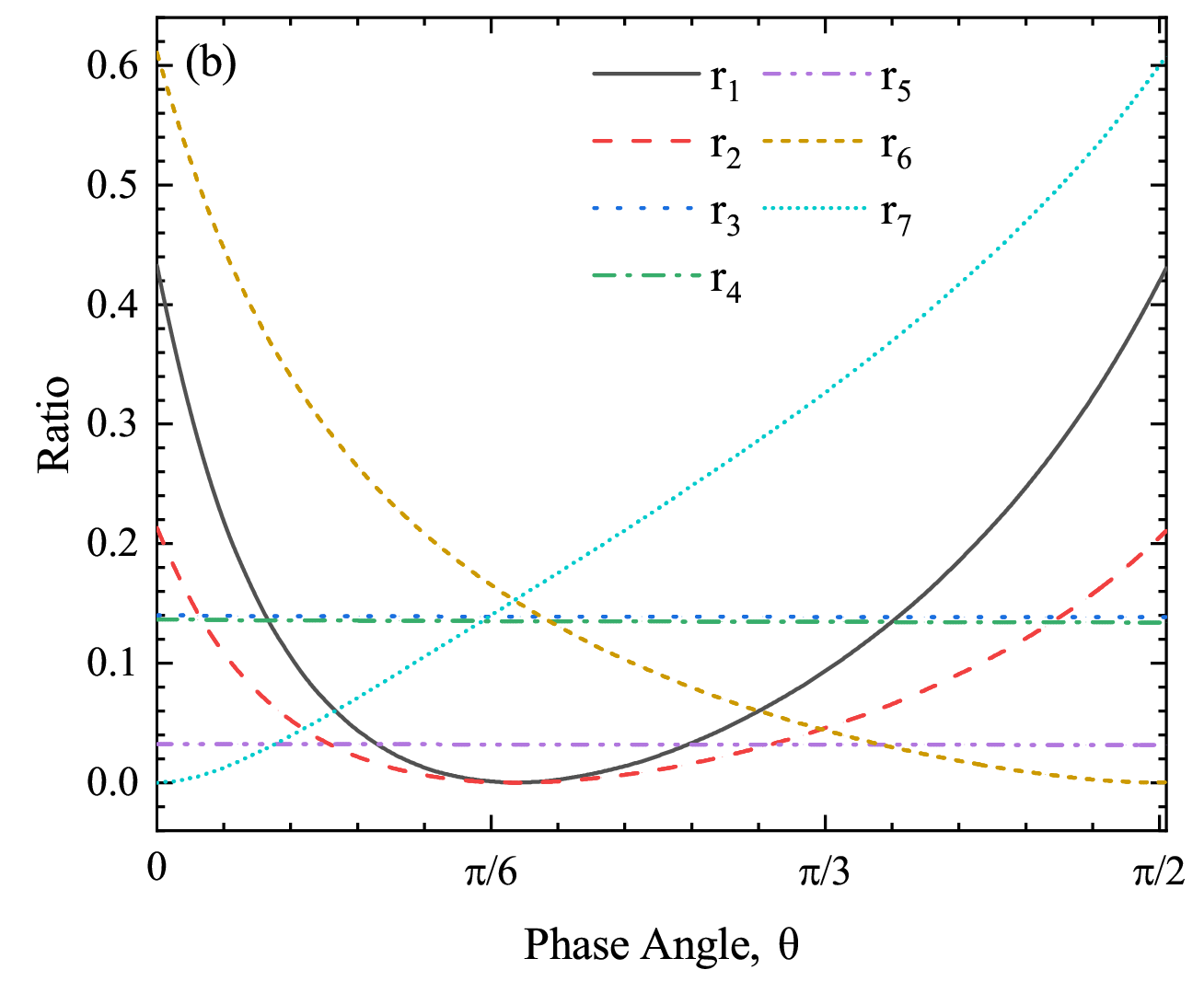}\,
	\caption{The ratios $R_i$ defined in Eq.~(\ref{eq:ratio_VV}) and $r_i$ defined in Eq.~(\ref{eq:ratio_PP}) as a function of the phase angle $\theta$. The model parameter $\alpha=1.0$ and the $\eta$-$\eta^\prime$ mixing angle $\theta_\mathrm{P} = -19.1^\circ$.}
	\label{fig:VV_PPtheta}
\end{figure*}

The ratios between different partial decay widths are used for studying the effects arising from the introduction of form factors. For the decays $X_2\to VV$, we define the following ratios with respect to the partial decay widths of $X_2\to \omega\omega$:
\begin{subequations}\label{eq:ratio_VV}
\begin{align}
    R_1 &= \frac{\mathrm{\Gamma}(X_2 \to \rho^0\omega)}{\mathrm{\Gamma}(X_2 \to \omega\omega)} \, , \label{eq:ratio_VV1}\\
    R_2 &= \frac{\mathrm{\Gamma}(X_2 \to \rho\rho)}{\mathrm{\Gamma}(X_2 \to \omega\omega)}\, , \\
    R_3 &= \frac{\mathrm{\Gamma}(X_2 \to K^{*+}K^{*-})}{\mathrm{\Gamma}(X_2 \to \omega\omega)} \, , \\
    R_4 &= \frac{\mathrm{\Gamma}(X_2 \to K^{*0} {\bar K}^{*0})}{\mathrm{\Gamma}(X_2 \to \omega\omega)} \,.
	\end{align}
\end{subequations}
Similarly, for the decays of $X_2 \to PP$, the following ratios are defined:
\begin{subequations}\label{eq:ratio_PP}
\begin{align}
    r_1 &= \frac{\mathrm{\Gamma}(X_2 \to \pi^0 \eta)}{\mathrm{\Gamma}(X_2 \to \pi \pi)} \, , \\
    r_2 &= \frac{\mathrm{\Gamma}(X_2 \to \pi^0 \eta')}{\mathrm{\Gamma}(X_2 \to \pi \pi)} \, , \\
    r_3 &= \frac{\mathrm{\Gamma}(X_2 \to \eta \eta)}{\mathrm{\Gamma}(X_2 \to \pi \pi)} \, , \\
    r_4 &= \frac{\mathrm{\Gamma}(X_2 \to \eta \eta')}{\mathrm{\Gamma}(X_2 \to \pi \pi)} \, , \\
    r_5 &= \frac{\mathrm{\Gamma}(X_2 \to \eta' \eta')}{\mathrm{\Gamma}(X_2 \to \pi \pi)} \, , \\
    r_6 &= \frac{\mathrm{\Gamma}(X_2 \to K^+ K^-)}{\mathrm{\Gamma}(X_2 \to \pi \pi)}\, ,\\
    r_7 &= \frac{\mathrm{\Gamma}(X_2 \to K^0 \bar{K}^0)}{\mathrm{\Gamma}(X_2 \to \pi \pi)}\,.
\end{align}
\end{subequations}

In Fig.~\ref{fig:VV_PPalpha}, we plot the ratios $R_1$ in terms of the model parameter $\alpha$. It indicates that the ratios are insensitive to the $\alpha$. From Fig.~\ref{fig:VV_PPalpha}, one can see that there is extremely strong dependence of the ratio on the phase angle $\theta$, which is of more fundamental significance than the parameter $\alpha$. This stability stimulates us to study the phase angle $\theta$ dependence.

In Fig.~\ref{fig:VV_PPtheta}, we plot the ratios $R_i$ ($i=1\sim4$) defined in Eq.~(\ref{eq:ratio_VV}) and $r_i$ ($i=1\sim7$) defined in Eq.~(\ref{eq:ratio_PP})  as a function of the phase angle $\theta$ with $\alpha=1.0$. One notice that the ratios $R_2$ and $r_{3,4,5}$ are independent of the phase angle $\theta$. These ratios shown in Fig.~\ref{fig:VV_PPtheta} may be tested by the future experimental measurements and can be used to determine the value of the phase angle.

\section{Summary}\label{sec:summary}

In this work, based on the assumption that the $X_2$ as a $D^*\bar{D}^*$ molecular state, we have investigated in detail the partial decay widths of $X_2 \to V V$ and $P P$ using the effective Lagrangian approach. For the $X_2$ state, we considered three cases, i.e., pure neutral components ($\theta = 0$), isospin singlet ($\theta = \pi/4$) and neutral components dominant ($\theta = \pi/6$), where $\theta$ is a phase angle describing the proportion of neutral ($D^{*0} \bar{D}^{*0}$) and charged ($D^{*+} D^{*-}$) constituents. When the $X_2$ is a pure neutral $D^{*0} \bar{D}^{*0}$ bound state, the predicted partial decay widths of the $X_2 \to V V$ and $X_2 \to P P$ are all several tens of $\mathrm{keV}$, corresponding to a branching ratio of $10^{-3}$--$10^{-2}$. However, when there are both neutral and charged components in the $X_2$, the decay widths of the isospin conserved processes $X_2 \to V V$ are predicted to reach several hundreds of $\mathrm{keV}$, leading to a upper limit of branching ratio of $10\%$, while the partial decay widths for the isospin conserved processes $X_2 \to P P$ are basically less than 100 keV, corresponding to a upper branching ratio of $10^{-2}$. For the isospin violated processes $X_2 \to V V$ and $X_2 \to P P$ are sensitive to the phase angle $\theta$. In the case of $\theta=\pi/6$, the width for the $X_2\to \rho^0\omega (\pi^0\eta^{(\prime)})$ is smaller than $1 (0.1)~\mathrm{keV}$, and it is between 3 (0.2) and 30 (5) keV for $\theta=\pi/4$.

We have also investigated the $X_2$ mass influence on the partial widths of the $X_2\to VV$ and $X_2\to PP$. The partial decay widths for the $X_2\to VV$ and $X_2\to PP$ exhibit similar behavior with varying the mass of $X_2$ state. Our results show that the partial decay widths for these processes are not very sensitive to the $X_2$ masses, unless the $X_2$ mass is quite closed to the $D^*\bar{D}^*$ threshold. Moreover, the dependence of these ratios between different charmless decay modes of $X_2$ to the charged and neutral phase angle for the $X_2$ in the molecular picture is also investigated, which may be tested by future experiments and can be used to determine the value of the phase angle.

\begin{acknowledgements}\label{sec:acknowledgements}

Zhao-Sai Jia thanks Zhen-Hua Zhang for useful discussions. This work is partly supported by the National Key R\&D Program of China under Grant No. 2023YFA1606703, and by the National Natural Science Foundation of China under Grant Nos. 12105153, 12075133, 12075288 and 12361141819, and by the Natural Science Foundation of Shandong Province under Grant Nos. ZR2021MA082, ZR2021ME147 and ZR2022ZD26. It is also supported by Taishan Scholar Project of Shandong Province (Grant No.tsqn202103062), the Higher Educational Youth Innovation Science and Technology Program Shandong Province (Grant No. 2020KJJ004), and the Youth Innovation Promotion Association CAS.

\end{acknowledgements}

\bibliography{X2decay_Refs.bib}
\end{document}